\documentclass[11pt,twoside]{article}

\usepackage{asp2014}

\aspSuppressVolSlug
\resetcounters

\begin{document}

\title{Integrating the VO Framework in the EOSC}

\author{Marco~Molinaro,$^1$ Mark~Allen,$^2$ Sara~Bertocco,$^1$ Catherine~Boisson,$^3$ Fran\c{c}ois~Bonnarel,$^2$ Margarida~Castro~Neves,$^4$ Markus~Demleitner,$^4$ Fran\c{c}oise~Genova,$^2$ Dave~Morris,$^5$ Andr\'{e}~Schaaff,$^2$ Giuliano~Taffoni,$^1$ and Stelios~Voutsinas$^5$}
\affil{$^1$INAF -- Osservatorio Astronomico di Trieste, Trieste, Italy; \email{marco.molinaro@inaf.it}}
\affil{$^2$Centre de Donn\'{e}es astronomiques de Strasbourg, Strasbourg, France}
\affil{$^3$Observatoire de Paris - Meudon, Paris, France}
\affil{$^4$Universit\"{a}t Heidelberg, Astronomisches Rechen-Institut, Heidelberg, Germany}
\affil{$^5$University of Edinburgh, Edinburgh, Scotland, UK}

\paperauthor{Marco~Molinaro}{marco.molinaro@inaf.it}{0000-0001-5028-6041}{INAF}{Astronomical Observatory of Trieste}{Trieste}{}{}{Italy}
\paperauthor{Mark~Allen}{mark.allen@astro.unistra.fr}{}{CNRS}{CDS}{Strasbourg}{}{}{France}
\paperauthor{Sara~Bertocco}{sara.bertocco@inaf.it}{}{INAF}{Astronomical Observatory of Trieste}{Trieste}{}{}{Italy}
\paperauthor{Catherine~Boisson}{catherine.boisson@obspm.fr}{}{CNRS}{Observatoire de Paris Meudon}{Paris}{}{}{France}
\paperauthor{Fran\c{c}ois~Bonnarel}{francois.bonnarel@astro.unistra.fr}{}{CNRS}{CDS}{Strasbourg}{}{}{France}
\paperauthor{Margarida~Castro~Neves}{mcneves@ari.uni-heidelberg.de}{0000-0001-9934-9234}{Universit\"{a}t Heidelberg}{Astronomisches Rechen-Institut}{Heidelberg}{}{}{Germany}
\paperauthor{Markus~Demleitner}{msdemlei@ari.uni-heidelberg.de}{}{Universit\"{a}t Heidelberg}{Astronomisches Rechen-Institut}{Heidelberg}{}{}{Germany}
\paperauthor{Fran\c{c}oise~Genova}{francoise.genova@astro.unistra.fr}{}{CNRS}{CDS}{Strasbourg}{}{}{France}
\paperauthor{Dave~Morris}{dmr@roe.ac.uk}{}{University of Edinburgh}{}{Edinburgh}{}{}{UK}
\paperauthor{Andr\'{e}~Schaaff}{andre.schaaff@astro.unistra.fr}{}{CNRS}{CDS}{Strasbourg}{}{}{France}
\paperauthor{Giuliano~Taffoni}{giuliano.taffoni@inaf.it}{0000-0002-4211-6816}{INAF}{Astronomical Observatory of Trieste}{Trieste}{}{}{Italy}
\paperauthor{Stelios~Voutsinas}{stv@roe.ac.uk}{}{University of Edinburgh}{}{Edinburgh}{}{}{UK}



\begin{abstract}
The European Open Science Cloud (EOSC) is in its early stages, but already some aspects of the EOSC vision are
starting to become reality, for example the EOSC portal and the development of metadata catalogues. 
In the astrophysical domain already exists an open approach to science data: 
the Virtual Observatory view put in place by the International Virtual Observatory Alliance 
(IVOA) architecture of standards.
The ESCAPE (European Science Cluster of Astronomy \& Particle physics ESFRI research 
infrastructures) 
project has, among its tasks, to demonstrate that the VO architecture can be integrated within the EOSC 
building one and to provide guidelines to ESFRI partners (European Strategy Forum on Research 
Infrastructures) in doing this.
This contribution reports on the progress of this integration after the first months of work inside ESCAPE.
\end{abstract}



\section{Introduction}
The European Open Science Cloud 
(EOSC\footnote{\url{https://ec.europa.eu/research/openscience/index.cfm?pg=open-science-cloud}}) 
is starting to address the needs of the large research infrastructures via the various projects that have been 
supported in H2020 programs. 
In the astrophysical domain it already exists an open approach to science data: 
the Virtual Observatory view put in place by the 
International Virtual Observatory Alliance (IVOA\footnote{\url{http://ivoa.net/}}).
The ESCAPE (European Science Cluster of Astronomy \& Particle physics ESFRI research 
infrastructures\footnote{\url{https://projectescape.eu/}}) 
project has, among its tasks, to demonstrate that the VO architecture can be integrated within the forming EOSC 
one and to provide guidelines to ESFRI partners (European Strategy Forum on Research 
Infrastructures) for EOSC integration of their own data holdings.
This contribution reports on the progress of this integration after the first months of work in the ESCAPE project.
After a quick overview of the EOSC goals (Sec.~\ref{p2.1:eosc}) and a brief description of the ESCAPE WP4 
Task-4.1 (Sec.~\ref{p2.1:cevo41}), the contribution summarises the planned and ongoing activities 
(Sec.~\ref{p2.1:planning}) and quickly describes the next steps (Sec.~\ref{p2.1:next}).

\section{The EOSC}
\label{p2.1:eosc}
Over the past years, numerous policy makers from around the world have articulated a clear and consistent 
vision of global Open Science as a driver for enabling a new paradigm of transparent, data-driven science. 
In Europe, this vision is being realised through the European Open Science Cloud (EOSC).
The EOSC will offer a virtual environment with open and seamless services for storage, management, 
analysis  and re-use of research data, across borders and scientific disciplines by federating existing 
scientific data infrastructures, currently dispersed across disciplines and the EU Member States.
Fig.~\ref{p2.1:fig:eoscroadmap} shows a representation of the EOSC Model action lines as available from the 
\textit{Implementation Roadmap for the European Open Science Cloud} document
\footnote{\hfill\raggedright\url{https://ec.europa.eu/research/openscience/pdf/swd_2018_83_f1_staff_working_paper_en.pdf}}.

\articlefigure{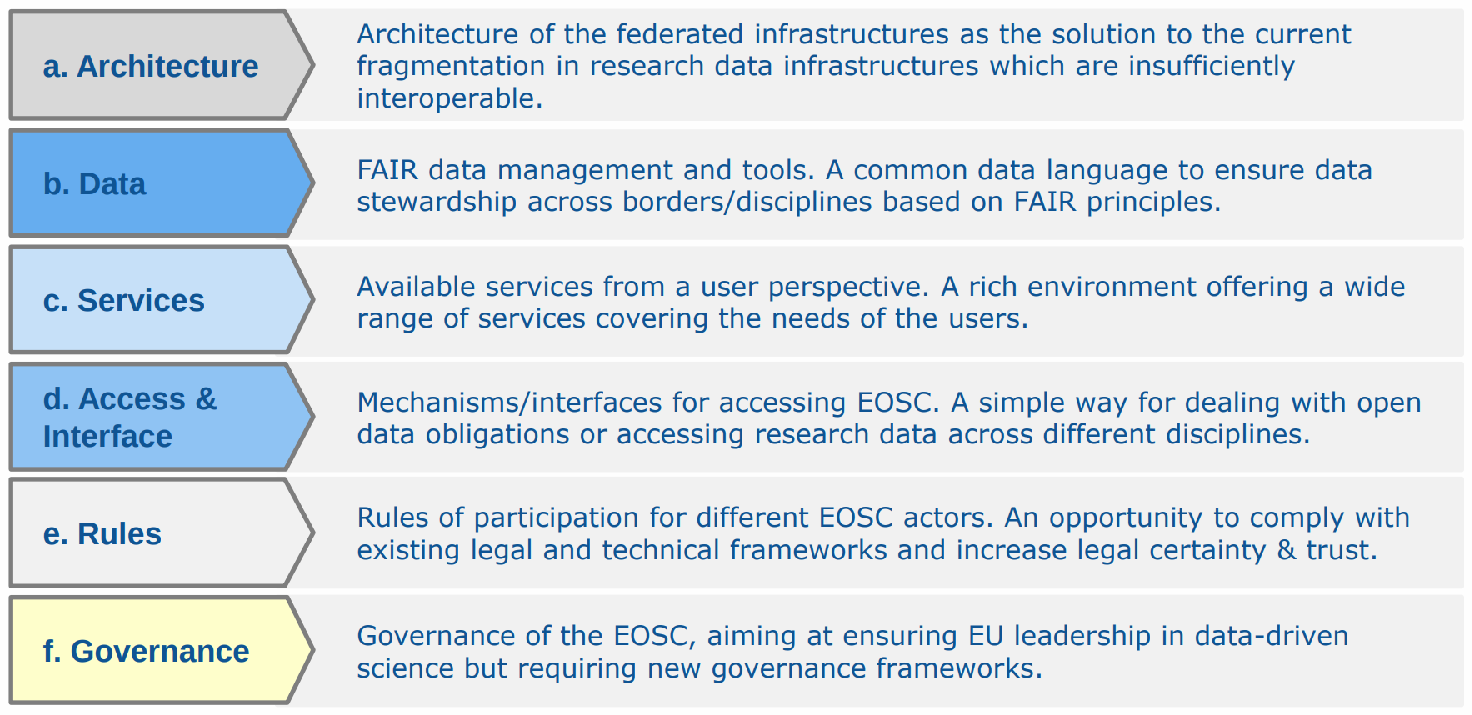}{p2.1:fig:eoscroadmap}{EOSC roadmap: action lines from the implementation 
document.}

\section{ESCAPE CEVO Task 4.1}
\label{p2.1:cevo41}
CEVO (Connecting ESFRI projects to EOSC through VO framework) is the work package 4 of the ESCAPE project.
It supports the continuing and substantial contribution to IVOA's activities in the framework of the EURO-VO (see, e.g., 
\citet{2015A&C....11..181G}) within the larger scenario the ESCAPE project is defining.
While refining the implementation of the FAIR principles \citep{2016NatSD...360018W} for astronomy data via 
common standards for 
interoperability and establishing data stewardship practices for adding value to the scientific content of 
ESFRI data archives, CEVO also aims, with Task 4.1, to map the VO framework to the EOSC, attempting to 
include the VO enabled archive services from ESFRIs in the latter. 

To reach its goals the task members will focus on four main points:
\begin{itemize}
    \item interfacing the VO framework with the EOSC;
    \item build an Astronomy portfolio of VO services;
    \item contributing to the EOSC Hybrid Cloud;
    \item containerising domain-specific services.
\end{itemize}

Planning and ongoing work for Task-4.1 is detailed in the nect section (Sec.~\ref{p2.1:planning}, while
Fig.~\ref{p2.1:fig:voeoscmap} shows a graphical representation of how the VO framework and technologies 
can be mapped onto the EOSC data lifecycle blocks (the diagram is taken from the ESCAPE's proposal 
Document of Work for the CEVO WP detailed description).

\articlefigure[width=0.8\textwidth]{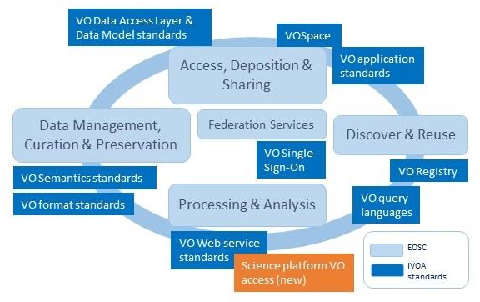}{p2.1:fig:voeoscmap}{VO mapping onto the EOSC data lifecycle.}

\section{Planned Activities for CEVO Task 4.1}
\label{p2.1:planning}

Activities for Task-4.1 have been planned among the Task participants and reported also in the first 
deliverable document of the CEVO WP. They can be roughly summarised as:
\begin{itemize}
    \item Inclusion of the VO Registry into the EOSC service catalogues;
    \item assessment of the methods for contributing an Astronomy Portfolio to the EOSC Marketplace;
    \item study of accessing VO-compliant data and services using science analysis platforms (in coordination
    with WP5);
    \item Assessment of the possibility to bring existing VO standards for data sharing (VOSpace) within the
    EOSC services;
    \item Identification of existing VO services or tools to serve as test-cases for containerisation.
\end{itemize}
In the first months, that included some ramp up and preparation of the work, a few points have been 
already worked out, namely:
\begin{itemize}
    \item based on the DataCite metadata schema\footnote{DataCite Metadata Schema: \url{https://schema.datacite.org/}}, metadata for the VO Registry's ~22000 services already 
    flows into EUDAT's B2-FIND\footnote{IVOA is listed as one of the \textit{communities} in it: \url{http://b2find.eudat.eu/group?q=ivoa}}; the translation from VOResource to DataCite 
    incurs a major loss
    of information, and we will work on identifying further kinds of VO metadata to represent in B2-FIND;
    \item contact points have been identified for catalogue integration and service portfolio;
    \item first steps in identifying a VOSpace back-end storage solution to allow VO standard data sharing;
    \item preliminary identification of the services and tools to be used in containerisation tests.
\end{itemize}

Among the latter, the most advanced activity is the one on the IVOA Registry Resources inclusion in the EOSC
catalogue.

\section{Further Steps}
\label{p2.1:next}

The deliverable document for the work plan of the CEVO package includes a list of activities to be brought
on. This will facilitate activity monitoring.

For Task-4.1 next steps to pursue include consolidating the contact points in the EOSC architecture to pave the way
to ESFRIs resources inclusion, inclusion of established VO services and technologies within the EOSC Hybrid Cloud 
and containerisation, but also bringing feedback to the EOSC building projects and stakeholders symposia.

\acknowledgements Authors acknowledge support from ESCAPE project (the European Science Cluster of 
Astronomy \& Particle Physics ESFRI Research Infrastructures) that has received funding from the European 
Union's Horizon 2020 research and innovation programme under the Grant Agreement n. 824064.

\bibliography{P2-1}  


\end{document}